\documentstyle[12pt,epsf,epsfig]{article}
\def\be{\begin{equation}}
\def\ee{\end{equation}}
\def\ba{\begin{eqnarray}}
\def\ea{\end{eqnarray}}
\def\la{~\mbox{\raisebox{-.6ex}{$\stackrel{<}{\sim}$}}~}
\def\ga{~\mbox{\raisebox{-.6ex}{$\stackrel{>}{\sim}$}}~}
\def\bq{\begin{quote}}
\def\eq{\end{quote}}

\newcommand{\beq}{\begin{equation}}
\newcommand{\eeq}{\end{equation}}
\newcommand{\beqa}{\begin{eqnarray}}
\newcommand{\eeqa}{\end{eqnarray}}

\newcommand{\lmk}{\left(}
\newcommand{\rmk}{\right)}
\newcommand{\lkk}{\left[}
\newcommand{\rkk}{\right]}
\newcommand{\lnk}{\left\{}
\newcommand{\rnk}{\right\}}
\newcommand{\Rbar}{\bar{R}}

\begin{document}

\thispagestyle{empty}
\begin{flushright}
CERN-TH/98-217\\ SU-ITP-98-44\\ UMN-TH-1710/98\\ YITP-98-41\\
hep-ph/9807482\\ July 1998
\end{flushright}
\vspace*{.5cm}
\begin{center}
{\Large \bf Topological $R^4$ Inflation}
 \\
\vspace*{.5cm}
John Ellis$^{a,}$\footnote{E-mail: john.ellis@cern.ch}, Nemanja
Kaloper$^{b,}$\footnote{E-mail: kaloper@epic.stanford.edu},
Keith A. Olive$^{c,}$\footnote{E-mail: olive@mnhep.hep.umn.edu} and
Jun'ichi Yokoyama$^{b,d,}$\footnote{E-mail:
yokoyama@yukawa.kyoto-u.ac.jp}\\
\vspace*{0.2cm}
$^{a}${\it Theory Division, CERN, CH~1211 Geneva~23, Switzerland}\\
\vspace*{0.2cm}
$^{b}${\it Department of Physics, Stanford University, Stanford, CA
94305-4060, USA}\\
\vspace*{0.2cm}
$^{c}${\it School of Physics and Astronomy, University of
Minnesota,}\\ {\it Minneapolis, MN 55455, USA}\\
\vspace*{0.2cm}
$^{d}${\it Yukawa Institute for Theoretical Physics, Kyoto
University, Kyoto 606-8502, Japan}\\
\vspace{1cm}
ABSTRACT
\end{center}
We consider the possibility that higher-curvature corrections
could drive inflation after the
compactification to four dimensions. Assuming that the
low-energy limit of the fundamental theory is 
eleven-dimensional supergravity to the lowest order, including
curvature corrections and taking the
descent from eleven dimensions to four via an intermediate
five-dimensional theory, as favored by recent considerations of
unification at some scale around $\sim 10^{16}$ GeV, we may obtain a
simple model of inflation in four dimensions. The effective
degrees of freedom are two scalar fields and the metric. The
scalars arise as the large five-dimensional modulus and the
self-interacting conformal mode of the metric. The effective potential 
has a local maximum in addition to the more usual minimum. However, the
potential is quite flat at the top, and admits topological inflation. We
show that the model can resolve cosmological problems and provide a
mechanism for structure formation with very little fine tuning.
\vfill

\setcounter{page}{0}
\setcounter{footnote}{0}
\newpage
\section{Introduction}

One of the central problems confronting inflation \cite{reviews}
is the identity of the inflaton, the field responsible for driving
inflation, and the manner in which it fits in with unified field
theories and/or string theory, notably $M$ theory. The birth of
the inflaton came with the demise of the notion that inflation is
driven by an adjoint Higgs field in some grand unified theory
(GUT) such as SU(5). While the production from quantum
fluctuations of the cosmological perturbations necessary to
generate structure in the universe is one of the great successes
of inflation \cite{pert}, the required smallness of the amplitude
of these fluctuations undermined the possibility that a field with
couplings of gauge strength could drive inflation. Rather, it is
often assumed that the inflaton couples to matter only through
gravitational interactions \cite{enot}. In this context, options
such as inflation via higher-dimensional curvature terms, including
as the $R^2$ inflation model proposed by Starobinsky \cite{star},
a hybrid inflationary model combining curvature and inflaton
effects, as discussed by Kofman, Linde and Starobinsky \cite{lk},
or string theory, which possesses many gauge-singlet fields such
as the dilaton \cite{bg,eenq}, may become quite interesting.

The most stringent constraints on inflation arise from the
observations of the cosmic microwave background. The naive
interpretation of the COBE and other data on fluctuations in the
microwave background radiation is that the density of vacuum
energy during inflation is $V \sim (10^{16}~{\rm GeV})^4$, so that
inflation is associated with an energy scale $V^{1/4} \sim
10^{16}$~GeV. One of the key points in the application of $M$ theory
to phenomenology is the reconciliation of the bottom-up
calculation of $M_{GUT} \sim 10^{16}$~GeV with the string
unification scale, which is close to the four-dimensional Planck
mass scale $M_4 \sim 10^{19}$~GeV. This is achieved by postulating
a large fifth dimension ${\cal R}_5 \gg M_{GUT}^{-1}$, which is
not felt by the gauge interactions, but causes the gravitational
interactions to rise with energy much faster than in the
conventional four dimensions. In this type of scenario, one could
expect that inflation should be considered within a five-dimensional
framework.

It is now known that ten-dimensional strongly-coupled heterotic
string theory is related through dualities to weakly-coupled type
I string theories, as well as to eleven-dimensional $M$
theory~\cite{m}. In each case, the corresponding thresholds imply
the presence of one large scale dimension below the unification
point \cite{m,bd}. Within this general five-dimensional framework,
two favored ranges for the magnitude of ${\cal R}_5$ can be
distinguished. One is relatively close to $M_{GUT}^{-1}$: ${\cal
R}_5^{-1} \sim 10^{12}$ to $10^{15}$~GeV, and the other could be
as low as ${\cal R}_5^{-1} \sim 1$~TeV \cite{aq}. The latter is
motivated in particular by Scherk-Schwarz models of supersymmetry
breaking, in which the gravitino mass $M_{3/2} \sim {\cal
R}_5^{-1}$. In this latter case, the large dimension is not
necessarily the conventional fifth dimension of $M$ theory.
Indeed, in models studied in~\cite{aq} the large dimension may be
related to what is normally considered as one of the six ``small"
dimensions that is conventionally compactified {\it \`a la} Calabi-Yau.

It has been known for quite
some time that it is very difficult to incorporate conventional
inflationary scenarios into the low-energy limit of string theory.
The principal obstacle in this course has been the fact that the
low-energy dynamics contains massless scalar fields with
non-minimal couplings to gravity. Their coupling constants are
precisely given by conformal symmetry and/or the dualities of
string theory. In an expanding universe, these fields typically
roll during the course of the expansion, consuming the available
energy and hence decreasing the rate of expansion. One typically
finds solutions where the scale factor of the universe grows as a
power of time, with the power determined by the scalar coupling
constants. Once the numerical values of these constants dictated by
string theory are taken into account, it has been found that the
resulting power laws are too slow to give an inflationary universe
\cite{clo}. Alternatively, if the scalars are endowed with masses
which arise from some kind of non-perturbative supersymmetry
breaking, the resulting models suffer from the graceful exit
problem, as we discuss below.
In string theory, higher-dimensional curvature terms are present in
the action, appearing in the expansion in powers of the string
tension. One might think that the inclusion of higher-derivative
terms, which are low-energy signatures of the massive excitations
of string theory, could produce several possibilities for
curvature-driven inflation.

In this work, we consider a variant of the Starobinsky model based
on $R^4$ curvature terms, as curvature-squared
terms are not known to be present in the action of the full five-
or eleven-dimensional theory. We assume throughout that the
remaining six dimensions are fixed~\footnote{For recent
discussions of fully eleven-dimensional cosmological solutions,
see~\cite{ikko,low}.}. In a four-dimensional context, Maeda
\cite{Maeda} derived the potentials for a general $R + R^n$
theory. For $n \ne 2$, the potential is not flat, but rather shows
a peak and a well-defined minimum. We argue that it is possible to
for this potential to inflate
in a so-called topological manner
\cite{topological}. As we discuss, a fully successful inflationary
model of this sort would still require a potential for the
dilaton. Whilst this model results in an inflationary stage with a
guaranteed exit, the magnitude and spectral index of the resulting
density fluctuations force us to consider some possible additional
ingredients. Either curvature-squared terms that might appear at
the level of the five-dimensional theory or quantum corrections to
$T_{\mu\nu}$~\footnote{The former can also be derived as a quantum
correction to the energy momentum tensor.} would be sufficient to
provide a self-consistent inflationary model.

Although such a
solution is not directly derived from string or $M$ theory, it
captures several elements that we expect to be present in the
low-energy theory. As such, it represents a novel and motivated
possible solution to the problem of inflation in string theory,
which may hold some promise.

\section{Curvature-Driven Inflation}

Among the first utilizations of higher-derivative curvature terms
is the Starobinsky model \cite{star}, which is based on obtaining a
self-consistent solution of Einstein's equations when they are
modified to include one-loop quantum corrections to the stress-energy
tensor $T_{\mu\nu}$. In its simplest form, the model is equivalent
to a theory of gravity with an $R^2$ correction. When one considers
the contributions of the back-reaction to the stress-energy, one
finds a term which is equivalent to starting with an action of the
form \cite{star2}:
\beq
\label{staract}
S={1 \over 2 \kappa^2}\int d^{4}x \sqrt{g} (R + R^2/6M^2)
\eeq
where $\kappa^2 = 8 \pi G_N$. It is well known that this theory is
conformally equivalent to a theory of Einstein gravity plus a
scalar field \cite{whitt}. By a field redefinition
\beq
{\tilde g}_{\mu\nu} = (1 + {1 \over 3 M^2} \phi) g_{\mu\nu}
~~~~~~~~~~
\phi' = \sqrt{{3 \over 2}}\ln (1 +  {1 \over 3 M^2} \phi)
\eeq
the action can be simplified to
\beq
\label{redact}
S={1 \over 2 \kappa^2} \int d^{4}x \sqrt{\tilde g} (R -
\partial^\mu \phi'
\partial_\mu \phi' - {3 \over 2} M^2 (1 - e^{-{\sqrt{2/3} \phi'}})
)
\eeq
The potential is extremely flat for $\phi' \gg M_4$ and has a
minimum at $\phi' = 0$ with $V(\phi' = 0) = 0$. For large initial
values of $\phi'$, one can recognize this as an excellent model for
chaotic inflation \cite{chaotic}.

More generally, quantum corrections to the right-hand side of
Einstein's equation in the absence of matter can be written as
\cite{corr}
\begin{eqnarray}
\langle T_{\mu\nu}\rangle & = & ({k_2 \over 2880 \pi^2 })(R_\mu^\rho R
_{\nu\rho}  - {2 \over 3} R R_{\mu\nu} - {1 \over 2}  g_{\mu\nu}
R^{\rho\sigma} R_{\rho\sigma} + {1 \over 4} g_{\mu\nu} R^2 )
\nonumber \\ & & + {1 \over 6}({k_3 \over 2880 \pi^2 })(2 R_{;\mu ;
\nu}    - 2g_{\mu\nu}   R^{;\rho}_{;\rho} - 2R R_{\mu\nu}  +  {1 \over
2} g_{\mu\nu} R^2 )
\label{tmncor}
\end{eqnarray}
where $k_2$ and $k_3$ are constants that appear in the process of
regularization. We recall that $k_2$ is related to the number of
light spin states, which can be very large in variants of string
theories based on $M$ theory, as we will discuss below. On the
other hand, the coefficient $k_3$ is independent of the number of
light states. This term is none other but the variation of the
$R^2$ term in the effective action. The theory admits a de Sitter
solution which can be found from the $00$ component of
gravitational equation of motion \cite{dc}. Defining $H' =
2880\pi^2
/ k_2$  and $M^2  = 2880\pi^2 / k_3$, and setting the spatial curvature
$k=0$, one finds \cite{vil}
\beq
H^2(H^2 - H'^2) = (H'^2/M^2)(2{\ddot H}H + 2H^2{\dot H} - {\dot H}^2 )
\label{h'}
\eeq
where $H = {\dot a}/a$ is the Hubble parameter. The de Sitter
solution corresponds to $H = H'$ and of course ${\dot H} = {\ddot
H} = 0$.

In order to avoid the overproduction
of gravitons there is a {\em lower} limit on the parameters
$k_{2,3}$ \cite{fp,vil}: $k_2 \ga 10^{10}$ implying the need for
billions of spin degrees of freedom to be present. While this seems
like an inordinately large number, it is possible to generate very large
numbers of degrees of freedom in theories with extra dimensions, as
we will now argue. 
Although this may not necessarily be the
framework for $M$ phenomenology that is eventually adopted, we use
the general form of the effective low-energy theory derived from
$M$-theory 
compactification on a Calabi-Yau manifold to illustrate
the discussion of the possibly large magnitude of $k_2$. 
The effective
low-energy field theory has the form of a five-dimensional
supergravity theory: as such, it contains a graviton
supermultiplet, vector supermultiplets and scalar hypermultiplets.
We recall that the graviton supermultiplet contains five graviton
states, three graviphoton states and eight gravitino states. Each
vector supermultiplet contains three vector states, one real scalar
and four fermion states, and each scalar hypermultiplet contains
four real scalars and four fermion states. The numbers of vector
hypermultiplets $n_V$ and scalar hypermultiplets $n_H$ are related
to the topological properties of the Calabi-Yau manifold:
\begin{equation}
n_V = n_{11} - 1, \; \; n_H = n_{21} + 1
\label{howmany}
\end{equation}
Some of these states have even parity when the fifth dimension is
compactified on $S_1/Z_2$, and some are odd, but this is not
essential for our purpose. We are interested in the number of
excited supermultiplets that appear below the effective
inflationary scale, which we identify approximately with
$10^{16}$~GeV~$\sim M_{GUT}$. The number of such Kaluza-Klein
excitations is given asymptotically by $n_{KK} \sim M_{GUT} {\cal
R}_5$. Hence we estimate
\begin{equation}
k_2 \sim n_{KK} \times \left( 16 + 8 (n_{11} + n_{21})
 \right)
\label{counting}
\end{equation}
In realistic models, we expect that $n_{11} = 3$ and $n_{21} =
n_{11} + 2 \chi$, where the Euler characteristic $\chi = 3$. In
this case, $k_2 \sim 88 n_{KK}$, which exceeds $10^{10}$ if
$n_{KK} \sim M_{GUT} {\cal R}_5 \ga 10^8$~\footnote{There is also
the possibility of additional matter and gauge fields
associated with $D$ branes in the bulk, but we do not discuss
them further here.}.

Although large, such a
value of $n_{KK}$ is perhaps not impossible as we shall see. In
the Starobinsky model, the bound for $k_3$ is $k_3 \ga  10^9$,
corresponding to $M \la 10^{14}$ GeV. This can be seen as follows.
In order to produce sufficient expansion to solve cosmological
problems, the effective cosmological constant, which by
(\ref{staract}) is $\sim M^2 M^2_4$, would have to be $M^2 M^2_4
\la 10^{64} ({\rm GeV})^4$. From this, we would find $M \la 10^{13}$
GeV. Alternatively, this is just the requirement that the Starobinsky
model embodies chaotic inflation, and simultaneously satisfies the
observational constraints. In our case, however, this requirement
can be relaxed since the energy for inflation will not be supplied
by the quadratic curvature term. As we will see below, our
inflation mass scale will still satisfy a similar inequality, but
without overconstraining $k_3$.

\section{Eleven-Dimensional Supergravity and Higher \\
Derivative Curvature Terms}

As we have indicated above, the presence of the dilaton field and
its universal coupling to other terms in the low-energy effective
action hamper the embedding of standard inflationary models in
string theory. One must remember that the derivative corrections
to the string effective action are not uniquely defined. Their
general form is fixed by requiring that the effective action
reproduces at two loops the $\beta$ functions of the string
world-sheet loop expansion \cite{corrs}. However there arise
divergences, and they must be renormalized. The results thus
depend on the particular subtraction scheme adopted in the
$\sigma$-model formulation. The renormalization-group
transformations relate different schemes, which changes the form
of the background $\sigma$-model couplings, while leaving the
physics invariant. These renormalization group transformations
viewed as maps of the target-space fields are called string field
redefinitions. They change the form of the effective action while
leaving the physics unaffected \cite{mt}.

If we return to the issue of the dilaton evolution, we see that in
general its equation of motion will be of the form $\nabla^2 \phi
- (\nabla \phi)^2 + R/4 \sim \alpha' J$, where $J$ is the
contribution of the higher derivative terms. It may contain higher
derivatives of $\phi$, such as $\nabla^4 \phi$ in addition to
curvature terms \cite{corrs,mt,mmjj}. These terms could make the
dilaton equation fourth-order in time derivatives. However, if we
take the limit $\alpha' \rightarrow 0$ in the dilaton equation,
for anything from $J$ to survive, it must diverge to cancel
$\alpha'$. Solutions of this type cannot be perturbatively matched
to the vacuum sector of the low-energy theory. Because of this one
cannot be certain that they will remain unaltered by higher-order
corrections. Also, the solutions are not uniquely determined at
the given order of truncation because of the string field
redefinitions discussed above. In terms of the dilaton field, this
suggests that the only physically meaningful effect of the source
$J$ at any given order of truncation is a perturbative correction
in $\alpha'$. This could be enforced at the level of the effective
theory by using field redefinitions to go to a unitary scheme,
where higher than second derivatives of fields are automatically
absent.

An example of such a unitary scheme is given by a four-dimensional
string gravitational action \cite{act} in the Einstein frame:
\begin{equation}
S = {1 \over 2 \kappa^2} \int d^4x \sqrt{g} \Bigl\{R
 - 2 \partial_{\mu} \phi
\partial^{\mu} \phi + \frac{\alpha'}{8} e^{-2 \phi}
\tilde R^2 \Bigr\}
\end{equation}
where ${\tilde R}^2 = R_{abcd} R^{abcd} - 4 R_{ab} R^{ab} + R^2$ is
the Gauss-Bonnet combination, and we have kept only the dilaton-
and metric-dependent terms. In the absence of a potential for the
dilaton, it has been shown that the combined dilaton and
gravitational equations of motion do not admit de Sitter solutions
\cite{nods,ko}. In fact, this result remains true when terms of
higher order in $\alpha'$ are considered \cite{bb}. As we show
below, this result also does not depend on the fact that it is the
Gauss-Bonnet combination. The coefficients of $R_{ab} R^{ab}$ and
$R^2$ are arbitrary up to a field redefinition of the metric and
dilaton \cite{mt}. For any choice of these coefficients, the only
solution with a constant dilaton is Minkowski space.

When a potential for the dilaton due to, e.g., gaugino condensation
\cite{gc} is included along with a cosmological term, due to,
e.g., a central charge deficit, it is possible to generate an
approximate de Sitter solution at the ${\cal O}(\alpha'^0)$ level
\cite{ko,nko1} when $\alpha'$ terms are kept. However, in this case
there is no exit from the inflationary period, and the dilaton is
already sitting at its minimum and is constant. We are led
therefore to a particular difficulty with field-theoretic inflation
in the gravitational sector of string theory~\footnote{We note in
passing a model which attempts to use running moduli to solve
cosmological problems: for the details of the model, see
\cite{pbb}, and for the discussion of its viability, see
\cite{ft}.}.

One can also choose to work in a non-unitary scheme, since all of
the schemes are physically equivalent. The spurious degrees of
freedom can be kept under control by expanding the source and
solving the equations iteratively. In such situations, the dilaton
would acquire ${\cal O}(\alpha')$ corrections as a response to the
source. While this rolling would appear to be adverse for
inflation, it is tempting to ask if it might only represent a
disguise. For example, it might happen that, by a field
redefinition, an apparently non-inflationary solution with a
rolling dilaton to ${\cal O}(\alpha')$ is mapped onto an
inflationary solution with a constant dilaton. This would require
retaining some of the spurious degrees of freedom, because the
constant dilaton Ansatz would demand cancellations between ${\cal
O}(1)$ and ${\cal O}(\alpha')$ terms. While perhaps unreliable,
such solutions are still of interest. They might be a starting
point for further study via more stringy methods.

Rather curiously, it turns out that imposing a constant dilaton in
an {\it arbitrary} subtraction scheme requires that the space-time
is exactly flat, and hence given by the Minkowski metric. To see
this, let us consider the effective action to ${\cal O}(\alpha')$.
In four dimensions, it can be written as
\be
\label{effact}
S = S_0 + S_1
\ee
where
\be
\label{effzero}
S_0 = \int d^4 x \sqrt{g} e^{-2\phi} \Bigl\{R + 4(\nabla \phi)^2
\Bigr\}
\ee
and
\ba
\label{effone}
S_1 &=& \alpha' \lambda_0 \int d^4x \sqrt{g} e^{-2\phi} \Bigl\{
R^2_{\mu\nu\lambda\sigma} + 2 \Bigl(R + 4\nabla^2 \phi - 4(\nabla
\phi)^2 \Bigr) \delta \phi
\nonumber\\
&& +\Bigl(R^{\mu\nu} + 2 \nabla^\mu \nabla^\nu - \frac12 g^{\mu\nu}
(R + 4 \nabla^2 \phi - 4 (\nabla \phi)^2 ) \Bigr) \delta g_{\mu\nu}
\Bigr\}
\ea
where we have chosen to work in the string frame, signified by the
presence of $\exp(-2\phi)$. The parameter $\lambda$ varies between
different string theories, being $1/4$ in bosonic, $1/8$ in
heterotic and $0$ in superstring theories. To ${\cal O}(\alpha')$,
only the square of the Riemann tensor is unambiguous. The
coefficients of all other terms are ambiguous, and can be set to
zero by redefining the fields in $S_0$ by terms to order $\alpha'$.
This is signified by the terms $\delta g_{\mu\nu}$ and $\delta
\phi$, which are
\ba
\label{deltas}
\delta \phi &=& c_1 R + c_2 (\nabla \phi)^2 + c_3 \nabla^2 \phi \nonumber \\
\delta g_{\mu\nu} &=& b_1 R_{\mu\nu} + b_2 \nabla_\mu \phi \nabla_\nu \phi
+ g_{\mu\nu} (b_3 R + b_4 (\nabla \phi)^2 + b_5 \nabla^2 \phi)
\ea
and are the most general expressions for the counterterms
consistent with dimensional analysis and target-space general
covariance. The coefficients $b_k$ and $c_k$ are completely
arbitrary.

Let us now consider the case when the coefficients $b_k$ and $c_k$
are chosen such that the source for the dilaton vanishes in a
cosmological Friedmann-Robertson-Walker (FRW) background. Due to the
dilaton equation of motion, this would place a constraint on the
curvature, which would select only those general metric solutions
of the dilaton-less theory that are a consistent truncation of the
${\cal O}(\alpha')$ string effective action. First, we see that if
the dilaton is a constant, all contributions to the action
proportional to $(\nabla \phi)^2$ would produce terms at least
$\sim \nabla \phi$ in the equations of motion, and so would vanish.
Inserting (\ref{deltas}) into (\ref{effone}), we see that the
${\cal O}(\alpha')$ action of interest to us is
\be
\label{effect}
S^{eff}_1 = \alpha' \lambda \int d^4x \sqrt{g} e^{-2\phi} \Bigl\{
R^2_{\mu\nu\lambda\rho} + b_1 R^2_{\mu\nu} + \beta_1 R^2 + \beta_2
\nabla^2 \phi R + 2 b_1 R_{\mu\nu} \nabla^\mu
\nabla^\nu \phi \Bigr \}
\ee
where, in terms of the original coefficients in (\ref{deltas}), we
have $\beta_1 = 2c_1 - b_3 - b_1/2$ and $\beta_2 = 8c_1 + 2c_3 -
b_5 - 6 b_3 - 2 b_1$.

If we now vary the action $S = S_0 +
S^{eff}_1$ with respect to $\phi$, and demand that $\phi = 0$ is a
solution, we obtain the following constraint on the curvature:
\be
\label{constr}
2 R + \alpha' \lambda \Bigl\{2R^2_{\mu\nu\lambda\rho} + 2b_1
R^2_{\mu\nu} + 2\beta_1 R^2
- (b_1 + \beta_2) \nabla^2 R \Bigr\} = 0
\ee
So when we consider the action $S_0 + S^{eff}_1$, we can set $\phi
= 0$ and impose the constraint (\ref{constr}) at the end. Further,
all FRW solutions are conformally flat. This means that the Riemann
curvature is given completely in terms of the Ricci tensor and
Ricci scalar. Simple algebra then shows that
\be
\label{riemann}
R^2_{\mu\nu\lambda\rho} = 2 R^2_{\mu\nu} - \frac13 R^2
\ee
The contributions of the Weyl tensor vanish because
$C^\mu{}_{\nu\mu\lambda} = 0$, which is true for any geometry, and
because on FRW backgrounds $C_{\mu\nu\lambda\rho}=0$. Thus, even
variations of the $C$-dependent terms would vanish on FRW
backgrounds. Since in four dimensions the Gauss-Bonnet term is
purely topological, i.e., is equal to a total divergence, we can
also write
\be
\label{gb}
R^2_{\mu\nu\lambda\rho} = 4R^2_{\mu\nu} - R^2 + \nabla_\mu J^\mu
\ee
for some vector field $J^\mu$ which is irrelevant for our
consideration. Combining the identities (\ref{riemann}) and
(\ref{gb}), we finds that on FRW backgrounds
\be
\label{ricci} R^2_{\mu\nu} = \frac13 R^2 + \frac12 \nabla_\mu
J^\mu \ee Since we are looking for solutions with $\phi = 0$, we
can drop the boundary term, as its variation would always be
proportional to $\nabla \phi$.

Using (\ref{riemann}) and
(\ref{ricci}), we find that the effective action on FRW
backgrounds with a constant dilaton is precisely the action of the
Starobinsky's model:
\be
S^{eff} = \int d^4x \sqrt{g} \Bigl\{R + \alpha' \lambda
(\frac{b_1}{3} + \beta_1 + \frac13) R^2 \Bigr\}
\ee
which must, however, be supplemented with the constraint
(\ref{constr}). If we vary this action with respect to the metric,
and trace the result, we get
\be
\label{trace}
R = \alpha' \lambda (6 \beta_1 + 2 b_1 + 2) \nabla^2 R
\ee
which is the equation of motion for the conformal mode of the
metric, that has become massive because of the derivative
corrections. In order for this equation to be consistent with the
constraint arising from requiring $\phi = 0$, we must ensure that
the terms in the dilaton constrain proportional to $R^2_{\mu\nu}$
and $R^2$ cancel identically. Using (\ref{riemann}), we find that
this requires setting $\beta_1 = 1/3$ and $b_1 = -2$. This
requirement is dictated by general covariance, and cannot be
relaxed to order $\alpha'$ if we wish to have a constant dilaton.
But when we insert this in the trace equation (\ref{trace}), we
find that the derivative term drops out, and we get
\be
\label{riczer} R = 0 \ee as a result. Worse yet, we see that the
coefficient of the $R^2$ term in the action is zero, and so the
vanishing of the dilaton requires that the metric is a solution of
the flat space equations to order $\alpha'$ and not only to order
$1$. Hence we see that the only solution of the ${\cal
O}(\alpha')$ action in any string theory with a constant dilaton
is Minkowski space, regardless of the subtraction scheme.

However, while some ways of compactifying the
eleven-dimensional theory to five dimensions are known \cite{m}, 
it is not yet clear if the compactification procedure is completely
unique. 
Discussion so far has centered on Calabi-Yau compactification,
whose features may depend on the mutual ratios of
sizes of the
interval $S^1/Z_2$ and $1$-cycles on the Calabi-Yau spaces, as has
already been pointed out in, for example,
\cite{aq}. We recall that there are alternative compactifications of the
weakly-coupled ten-dimensional heterotic string,
and they may turn
out to have elevations to the eleven-dimensional theory.
In this article, we therefore
consider a general approach in which we 
assume the internal manifold to be decoupled,
with its size a massive field from the point of view of the
five-dimensional theory. There have been several
calculations of higher-order $R^n$ terms in ten and eleven
dimensions \cite{R4}. Those in eleven dimensions are not known to
yield $R^2$ terms, but may yield $R^4$ terms. On the other hand,
the calculations by Ho\v rava and Witten on the ten-dimensional
boundary in their formulation of $M$ theory do yield $R^2$ terms,
but only in the boundary effective actions.
We will not discuss their precise form, 
which in five dimensions may depend
on the details of compactification of the eleven-dimensional theory. If we
dimensionally reduce the five-dimensional theory to four
dimensions, we will obtain an even more complicated-looking
expression involving contractions of the four-dimensional Riemann
tensor and terms with up to two derivatives of the size of the
internal dimension, in addition to the four-dimensional version of
the five-dimensional expression. Now, as long as the four-dimensional
space-time is conformally flat, its Riemann tensor can be
expressed completely in terms of the Ricci tensor and scalar, and
thus any quartic curvature term could be written as a linear
combination of terms like $R^4, R^2_{\mu\nu}$ and $R_{\mu\nu}
R_{\lambda\sigma} R^{\mu\lambda} R^{\nu\sigma}$. Moreover, in four
dimensions, the Gauss-Bonnet identity allows us to replace the
square of the Ricci tensor by a square of the Ricci scalar,
indicating that the scalar curvature modes play the most important
role in cosmological dynamics. Here we repeat that the scalar
modes must be endowed with mass in order that they decouple. This
means that the terms proportional to its derivatives would all
drop out, and hence we will ignore the scalar-tensor couplings
which must depend on the derivatives of the scalar fields.

To model the possibility of curvature-driven inflation, we assume
that the five-dimensional action contains a scalar $R^4$
contribution, and perform a dimensional reduction to four
dimensions, ignoring the boundaries. In the context of the theory
with walls, this merely means that we assume that the bulk can be
foliated by identical and mutually non-interacting copies of the
wall. Alternatively, it is clear that this dimensional reduction is
identical to the standard Kaluza-Klein reduction on a circle. This
approach produces a four-dimensional theory with two scalar fields,
an inflaton and a dilaton. The inflaton potential has a maximum,
which we will show to be sufficiently flat to support inflation, in
a manner resembling the topological inflationary scenario. A
problem with the scenario based only on the $R^4$ term, without the
mass term for the compacton, is that it wants to decompactify the
space time. This is obvious if we consider the $R^4$ term from the
point of view of five dimensions. Since it behaves as an effective
cosmological constant, according to the cosmological no-hair
theorem, it forces the five-dimensional space to isotropize. Thus
we must include a mass term for the compacton, which from the
five-dimensional point of view will break the rotation symmetry,
and allow the four-dimensional space to inflate while keeping the
compacton fixed.

\section{Inflation In Four Dimensions}

In this section, we will begin the investigation of the possibility
of higher-curvature-driven inflation. We first give details of the
dimensional reduction from five dimensions to four dimensions,
in order to derive the
four-dimensional effective theory. The pure gravity sector of the
five-dimensional action, which we will consider is
\be
\label{act5} S= \int d^{5}x \sqrt{G_5} \Bigl\{{M_5^3 \over 16 \pi}
R_5 + \alpha M_5^{-3} R^4_5 \Bigr\} \label{f5d} \ee where $ M_5$
is the five-dimensional Planck mass. In the context of the
eleven-dimensional theory, our
five-dimensional Planck mass is related to the
eleven-dimensional Planck mass by $M_5^3 =
M_{11}^9 V_6$, where $V_6$ is the six-volume of the compactified
space. We assume that $V_6^{-1/6} \sim M_{11} \sim M_{\rm GUT}$ in
general. The scale of the fifth dimension is then given by ${\cal
R}_5^{-1} \sim {4 \over \alpha} M_{11}^3 M_4^{-2}$ \cite{aq}.
We note that this effective action is not exactly what one
finds after dimensional reduction of the eleven-dimensional supergravity
with
quartic corrections on an interval $S^1/Z_2$, which would contain
couplings
to two ten-dimensional boundaries with $R^2$ terms (see, e.g.,
\cite{low}). 
Thus,
our  action (\ref{act5}) is not a direct descendant of a known
reduction of
$M$ theory on an interval. But such actions may nevertheless arise in some
compactifications of the theory, and are simple enough to illustrate our
main point.

Let us first outline the reduction procedure we will follow here.
We first conformally transform the action (\ref{act5}), following
\cite{Maeda}, to represent it as a five-dimensional gravity with a
minimally
coupled self-interacting scalar field. The scalar arises because
the conformal mode of the metric becomes dynamical thanks to the
$R^4$ term \cite{whitt}. Then we dimensionally reduce this action
to four dimensions and apply another, four-dimensional, conformal
transformation, to put the resulting four-dimensional action in the
canonical
form. This will produce another scalar field, the compacton (or
dilaton) of the four-dimensional theory, which is related to the size of
the
fifth dimension.

We now give the formulae appropriate for this procedure. The
conformal transformation which brings (\ref{act5}) to the Einstein
form is
\beq
 \bar{G}_{AB}=| 1+64\pi\alpha M_5^{-6} R_5^3|^{\frac{2}{3}}G_{AB},
\eeq
and the action is \cite{Maeda}
\newcommand{\chit}{\tilde{\chi}}
\newcommand{\kappat}{\tilde{\kappa}}
\beq \label{act5bar} \bar{S}_5=\int d^{5}x \sqrt{\bar{G}_5} \lnk
\frac{M_5^3}{16\pi}\Rbar_5-\frac{1}{2}
\bar{G}^{AB}\partial_A\chit\partial_B\chit-U_5(\chit)\rnk. \eeq
where the ``intermediate" scalar field and its potential are \beq
\kappat\chit\equiv \frac{2}{\sqrt{3}}\ln|1+64\pi\alpha
M_5^{-6}R_5^3|,~~~~~~
U_5(\chit)=\frac{3M_5^5}{256\pi^{\frac{4}{3}}\alpha^{\frac{1}{3}}}
e^{-\frac{5\sqrt{3}}{6}\kappat\chit}\lmk
e^{\frac{\sqrt{3}}{2}\kappat\chit}-1\rmk^{\frac{4}{3}}. \eeq with
$\kappat\equiv \sqrt{(8\pi)/(M_5^{3})}$. The standard Kaluza-Klein
compactification Ansatz, with the simultaneous conformal
transformation of the
four-dimensional metric to the canonical form, is \beq
\label{kkans} d\bar{s}_5^2= e^{-\sqrt{\frac23} \kappa \phi}
g_{\mu\nu}dx^\mu dx^\nu + \frac{M^4_4}{M^6_5} e^{2\sqrt{\frac23}
\kappa \phi} dz^2, \eeq where $g_{\mu\nu}$ is the four-dimensional
Einstein-frame metric. The action (\ref{act5bar}) then reduces to
the following four-dimensional action: \beq \label{act4} S_4=\int
d^4x\sqrt{g_4}\lnk {M_4^2 \over 16 \pi} R_4
-\frac{1}{2}g^{\mu\nu}\partial_\mu\varphi\partial_\nu\varphi
-\frac{1}{2}g^{\mu\nu}\partial_\mu\chi\partial_\nu\chi
-U(\varphi,\chi)\rnk. \eeq where
$\chi=M_4M_5^{-\frac{3}{2}}\chit$.
In terms of these fields, the
potential is given by \beq \label{potential}
U(\varphi,\chi)=\frac{3 M_4^2
M_5^2}{256\pi^{\frac{4}{3}}\alpha^{\frac{1}{3}}}
e^{-\sqrt{\frac{2}{3}} \kappa \varphi} e^{-\frac{5 \sqrt{3}}{6}
\kappa \chi} \lmk e^{\frac{\sqrt{3}}{2} \kappa
\chi}-1\rmk^{\frac{4}{3}} \equiv \frac{3 M_4^2
M_5^2}{256\pi^{\frac{4}{3}}\alpha^{\frac{1}{3}}}
e^{-\sqrt{\frac{2}{3}} \kappa \varphi}V(\chi). \label{upot}
\eeq
Before we undertake a detailed investigation of (\ref{act4}), we
note that the effective
four-dimensional potential (\ref{upot}) is a product
of an exponential and a function with a maximum. If we ignore the
variation of $\chi$, we recall that in a universe dominated by a
scalar field with an exponential potential $V(\phi)\propto
e^{-\lambda\kappa\phi}$, and foliated by flat spatial hyperplanes,
the expansion of the FRW scale factor $a(t)$ obeys a power law
$a(t) \propto t^p$, with the power index given by
$p=2/\lambda^{2}$ \cite{pli}.

\begin{figure}[h]
\begin{center}
\epsfig{file=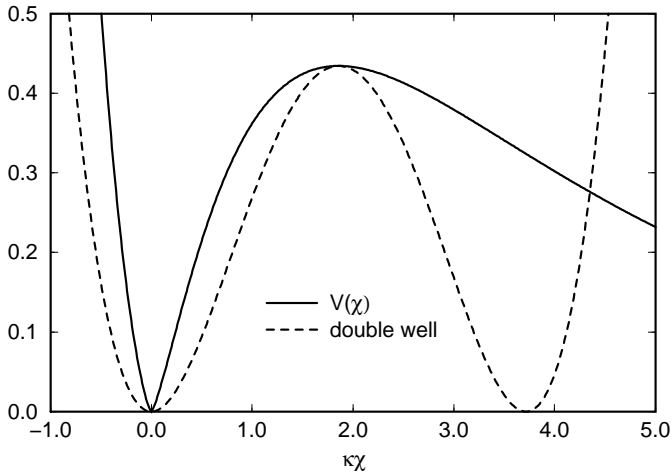,height=9cm}
\caption{\label{fig:fig1} {\it Comparison of the potential $V(\chi)$
produced by a conformal rescaling of the $R^4$ term in the
five-dimensional action and a
simple double-well potential.}} \ \end{center}
\end{figure}

Next, we ignore the dilaton factor $e^{-\sqrt{\frac{2}{3}} \kappa
\varphi}$ in the potential and consider the dynamics of the $\chi$
field. The $\chi$-dependent factor in the potential, $V(\chi)$, is
depicted in Fig.~1. It has a global minimum $V=0$ at $\chi=0$, a
local maximum at $\chi=\chi_m= 2\ln 5/(\sqrt{3}\kappa)=0.37M_4$,
and diverges as $V(\chi)\propto e^{-\frac{5\sqrt{3}}{6}\kappa\chi}$
for $\chi\longrightarrow -\infty$. This is too steep for power-law
inflation. For $\chi\longrightarrow\infty$, $V(\chi)$
asymptotically approaches zero with $V(\chi)\propto
e^{-\frac{\sqrt{3}}{6}\kappa\chi}$, which may be flat enough for
inflation but it will lead to an unphysical universe with a runaway
behavior towards the regime of extremely large curvature, in terms
of the original five-dimensional description.

As a result, chaotic inflation \cite{chaotic} with a large
initial value of $\chi$ is impossible here. Nevertheless there may
remain the possibility to realize inflationary expansion in this
model by using the potential energy around the local maximum,
$V(\chi_m)$, as in the topological inflation scenario of Linde and
Vilenkin \cite{topological}. In this scenario, if the scalar field
$\chi(x)$ is randomly distributed initially with a large
dispersion, some part of the universe will roll to $\chi=0$, while
in other parts it will run away to infinity. Between any two such
regions there will appear domain walls, containing a large energy
density, $\rho \sim V(\chi_m)$. If the wall is thicker than the
Hubble radius of this energy density, there will exist a
sufficiently large quasi-homogeneous region, filled with large
potential energy, where inflationary expansion naturally sets in.

The condition for a domain wall to inflate has been
investigated numerically
in~\cite{Sakai} for the case of a simple double-well
potential, $V_{dw}(\phi)=(\frac{\lambda}{4})(\phi^2-\eta^2)^2$. There it
has been found that a domain wall will undergo inflation if $\eta$
exceeds a critical value $\eta_{cr}=0.33M_4$, regardless of the
value of $\lambda$. When $\eta=\eta_{cr}$, the ratio of the
thickness of the wall - characterized by the curvature scale of the
potential at the origin, $r_w\equiv \lmk
V_{dw}''(\phi=0)\rmk^{-\frac{1}{2}}$, to the horizon
$H^{-1} = \lmk \kappa^2V_{dw}(\phi=0)/3\rmk^{-\frac{1}{2}}$ is
given by $r_wH=0.48$, and is again independent of $\lambda$. An
explicit check shows that, in our model, the distance between the
potential minimum and the local maximum, $\chi_m=0.37M_4$, exceeds
$\eta_{cr}$. Furthermore, the ratio of the characteristic thickness
of the wall to the horizon scale is given by $r_wH=\lmk
\kappa^2V(\chi_m)/3V''(\chi_m)\rmk^{\frac{1}{2}}=4/\sqrt{15}
\simeq 1.0$, which
is larger than the critical case of~\cite{Sakai}. In Fig.~1 we have
also depicted a double-well potential which has the same global
minimum and the local maximum as $V(\chi)$. As is seen there, the
latter is much flatter than the former around the local maximum.

Having seen that the potential $V(\chi)$ is flat enough around the
local maximum, we now return to the exact form of the
potential given in (\ref{upot}), and take the dilaton factor into
account. This does not change $\chi_m$ nor $r_wH$. However, the
rolling dilaton field hampers exponential inflation. Since $\chi$
moves slowly near $\chi=\chi_m$, as compared to the dilaton, we
practically have an exponential potential \beq U(\varphi,\chi)=
\frac{3 M_4^2 M_5^2}{256\pi^{\frac{4}{3}}\alpha^{\frac{1}{3}}}
e^{-\sqrt{\frac{2}{3}} \kappa \varphi}V(\chi_m)\equiv
U_0e^{-\sqrt{\frac{2}{3}} \kappa \varphi}. \eeq The ensuing
solution for the FRW scale factor obeys the power law $a(t)\propto
t^3$ with $\kappa\varphi(t)=\kappa\varphi(t_{i})
+\sqrt{6}\ln(t/t_{i})$, and is an asymptotic attractor
\cite{attractor} for the scale factor. Since the power is greater
than unity, this solution still appears to be slowly inflating.
However, when we look at the scalar field, we find
$\exp(\sqrt{2/3} \kappa \varphi) = t^2(\pi U_0)/(3M_4^2)$. This
evolution of the scalar field is too rapid, as we can easily
verify by substituting the solutions back into the
five-dimensional metric
(\ref{kkans}). Indeed, in the original frame the solution is
$\bar{a}(\bar{t})\propto \exp\lmk\sqrt{(4\pi
U_0)/(3M_5^3)}\bar{t}\rmk$, and $\exp(\sqrt{2/3} \kappa \varphi)
\propto \exp\lmk\sqrt{(4\pi U_0)/(3M_5^3)}\bar{t}\rmk$. The
physical interpretation of this behavior is simple. Substituting
$\chi=\chi_m$ in the potential corresponds to adding a positive
cosmological constant in five dimensions, and, according to Wald's
cosmological no-hair theorem \cite{Wald}, the anisotropic
five-dimensional spacetime we are dealing with must approach the
de Sitter space due to the effective cosmological constant.

Hence some mechanism is needed to stabilize the size of the fifth
dimension. This requires breaking the residual
five-dimensional gauge
invariance, which equates the fifth direction with the other four
in the case discussed above. 
The problem is linked to that of fixing the v.e.v. of the dilaton,
which presumably involves ill-understood non-perturbative phenomena
such as supersymmetry breaking and perhaps gaugino condensation
\cite{gc}. Scenarios for these have been proposed, and it seems
quite possible that these may occur at an energy-momentum scale
above that of the five-dimensional Kaluza-Klein excitations. In
this case, the internal radius may be regarded as fixed, for our
purposes here.

After the internal radius is stabilized, we can recover exponential
inflation in three spatial dimensions in both conformal frames, as
long as $\chi$ stays near the local maximum $\chi_m$. In this
limit, the potential can be approximated as
\ba
U(0,\chi)= \frac{3 M_4^2
M_5^2}{256\pi^{\frac{4}{3}}\alpha^{\frac{1}{3}}} V(\chi) \sim
U(0,\chi_m)-\frac{1}{2}m^2(\chi-\chi_m)^2,~~~
m^2\equiv \frac{5}{16}\kappa^2U(0,\chi_m),
\ea
where we have used the relation $V''(\chi_m)=-\frac{5}{16}\kappa^2
V(\chi_m)$. Thus the standard slow-roll solution is
\beqa
a(t)\!\! &\propto & \!\!e^{H_m t},~~~~~H_m\equiv
\sqrt{\frac{\kappa^2}{3}U(0,\chi_m)},  \\
\chi(t)\!\! &=&\!\! \chi_m-(\chi_m-\chi_f)\exp\lmk
\frac{m^2}{3H_m^2}H_m(t-t_f)\rmk = \chi_m-(\chi_m-\chi_f)
e^{\frac{5}{16}H_m(t-t_f)}, \nonumber
\eeqa
where $t<t_f$, and where $t_f$ and $\chi_f$ stand for the time and
the field amplitude at the end of inflation.

It is important to note that the model we are presently
considering does not share the graceful exit problem of typical
string-dilaton inflationary models. Generally, as we have described
above, inflationary (or de Sitter) solutions with a dilaton coupled
to gravity require that the dilaton be fixed. When the dilaton
doubles as the inflaton as well, there no means have been found to
cancel the vacuum energy which drives inflation. This is different
from the graceful exit problem in the pre-big-bang scenarios for
which inflation is not driven by vacuum energy density, though
these models also suffer from a graceful exit problem \cite{ge}. In
our present case, these two issues are separated. Although the
dilaton still needs to be fixed, inflation is driven by the
$R^4$-induced potential of the conformal field $\chi$. Since the
model we are considering is of the ``topological" type, we are
sitting on the top of this potential and we are {\em guaranteed}
that inflation will end as the field $\chi$ rolls to its minimum.

We can now check the magnitude and spectral index of the induced
density fluctuations. The amplitude of a linear curvature
fluctuation, $\Phi_H$, \cite{Bardeen} on a comoving scale $l=2\pi/k$
is given by
\beqa
\Phi_H\lmk l=\frac{2\pi}{k}\rmk = \frac{fH_m^2}{2\pi|\dot{\chi}(t_k)|}
&=&\frac{3}{2\pi}\frac{f H_m^3}{m^2 |\chi_m -\chi_f|}
\exp\lmk\frac{m^2}{3H_m^2}H_m(t_f-t_k)\rmk \\
&=&\frac{8fH_m}{5\pi |\chi_m-\chi_f|}e^{\frac{5}{16}H_m(t_f -
t_k)}, \nonumber
\eeqa
where $f=3/5~(2/3)$ in the matter- (radiation-) dominated era, and
$t_k$ is the time when the $k$ mode left the Hubble radius during
inflation. The spectral index, $n_s$, of density perturbation is
given by
\beq
n_s=1-\frac{2m^2}{3H^2_m}.
\eeq
We recall that, in the model we are discussing, $m^2/H_m^2 = 15/16$
so that $n_s = 3/8$, which is significantly different from the
scale-invariant value $n_s=1$ and is in disagreement with
observations. Furthermore, the large-angle (an)isotropy of cosmic
microwave background radiation (CMB) \cite{COBE} requires $\delta
T/T=-\Phi_H/3=10^{-5}$ on the comoving scale leaving the Hubble
radius about 60 expansion times before the end of inflation,
namely,
\beq
\frac{\delta T}{T}=\frac{8H_m}{25\pi
|\chi_m-\chi_f|}e^{\frac{5}{16}\times 60}\simeq 10^{-5}.
\eeq
Since we find $|\chi_m-\chi_f|\sim 0.1M_4$, the isotropy of CMB
sets the scale of inflation as $H_m\sim 10^{-14}M_4$, which implies
the Planck mass in five dimensions must be unacceptably small:
$M_5\sim 10^{-13}\alpha^{\frac{1}{6}}M_4$.

These results appear disconcerting. While our model showed some
initial promise, it seems to fail the contact with the
observations. A closer scrutiny of the dynamics shows that these
problems arise because the potential which is generated by the
$R^4$ term is a bit too steep to produce a satisfactory
perturbation spectrum. In the next section, we discuss
possible remedies.

\section{Cures For The Density Fluctuation Problem}

As we have seen in the previous section, although the relatively
simple model described there, which is based on a compactified $R +
R^4$ theory with a fixed dilaton, has all the ingredients necessary
for inflation, we cannot obtain an acceptable magnitude for density
fluctuations unless we choose the scale $M_5 \simeq 10^{-14} M_4$.
But if on the other hand we have $M_5 \sim M_{11}
\sim M_{\rm GUT}$, as occurs when $V_6^{-1/6} \sim M_{11}$ with $M_5^3
\sim M_{11}^9 V_6$, then by setting
$M_5 \sim M_{\rm GUT} \sim 10^{-3}
M_4$ we would have greatly overproduced the magnitude of density
fluctuations in the model. Moreover, we would still have the
problem that the spectral index is equal to $3/8$. We
could foresee two possible solutions to this dilemma. The presence
of an $R^2$ term in the five-dimensional action would flatten the
potential further, possibly curing the problems with density
fluctuations. In addition, the renormalization of the stress energy
tensor, as in the original Starobinsky model, could also lower the
effective value of $m^2/H_m^2$, which was at the root of the
problems above.

The general classification of higher-order $R^n$ terms which may appear in the
curvature expansion of the effective low-energy field-theory limit of $M$ 
theory is not available at present. It is however known that
the supersymmetry of the theory rules out terms which are quadratic and
cubic in curvature in the bulk, and that the lowest possible terms are
quartic. Upon dimensional reduction of the eleven-dimensional theory on
the interval $S^1/Z_2$ {\it \`a la}  Ho\v rava-Witten, such terms could
however produce terms which are quadratic in curvature in the effective
action of the boundary theory. Our considerations here are
different, since we do not consider theories with matter degrees of
freedom confined only to the boundary. We should note that
higher-order formulations of higher-dimensional gravity and
supergravity theories  have been discussed in~\cite{h}. 
Whilst the known types of quadratic corrections arising
from $M$ theory are not explicitly of the type we need here,
their form being restricted by supersymmetry, 
we recall that such
constraints are relaxed
in cases when supersymmetry is broken. Moreover, 
if we consider the effect of particle production and its back-reaction
on the geometrical environment, we recall that this effect could be 
derived from effective counterterms in the
action which are quadratic in curvature.

Therefore, we boldly consider the case where the
five-dimensional action
contains both quadratic and quartic terms in $R_5$
\beq
S_5=\int d^{5}x \sqrt{G_5} \Bigl\{ \frac{M_5^3}{16\pi}R_5 +b
M_5R_5^2 + c M_5^{-3}R^4_5\Bigr\}.
\label{5dqq}
\eeq
As we noted above, because of the absense of boundary terms, this action
is not directly related to the Ho\v rava-Witten reduction of $M$-theory.
We can again carry out the reduction to four dimensions along the same
lines as discussed at the beginning of the previous section. Assuming
$b,~c >0$, we apply the conformal transformation
\beq
\bar{G}_{AB}=
| 1+32\pi bM_5^{-2}R_5 +64\pi c M_5^{-6} R_5^3|^{\frac{2}{3}}G_{AB},
\eeq
to obtain the equivalent action in the Einstein frame \cite{Maeda}:
\newcommand{\chih}{\hat{\chi}}
\beq
\label{act5barqq}
\bar{S}_5=\int d^{5}x \sqrt{\bar{G}_5}
\lnk \frac{M_5^3}{16\pi}\Rbar_5-\frac{1}{2}
\bar{G}^{AB}\partial_A\chih\partial_B\chih-\hat{U}_5(\chih)\rnk,
\eeq
with
\beq
\kappat\chih\equiv \frac{2}{\sqrt{3}}\ln|1
 +32\pi b M_5^{-2}R_5+64\pi c M_5^{-6}R_5^3|. \label{chih}
\eeq Since we are assuming $b,~c>0$, there is a one-to-one
correspondence between $\chih$ and $R_5$ for $1 +32\pi b
M_5^{-2}R_5+64\pi c M_5^{-6}R_5^3 > 0$, and from (\ref{chih}) we
can solve for $\chih$ to find the potential
\beq
\hat{U}_5(\chih)=M_5^5 e^{-\frac{5\sqrt{3}}{6}\kappat\chih} \lmk
-\frac{bR_5^2}{2M_5^4} +2u(\chih)\frac{R_5}{M_5^2}\rmk,~~~
u(\chih)\equiv \frac{e^{\frac{\sqrt{3}}{2}\kappat\chih}-1}{128\pi c},
\eeq
with
\beq
 R_5=M_5^2\lnk u(\chih)+\lkk u^2(\chih)+\lmk\frac{b}{6c}\rmk^3
 \rkk^{\frac{1}{2}}\rnk^{\frac{1}{3}}-
 \lnk -u(\chih)+\lkk u^2(\chih)+\lmk\frac{b}{6c}\rmk^3
 \rkk^{\frac{1}{2}}\rnk^{\frac{1}{3}}.
\eeq
We compactify to four dimensions and
apply another conformal transformation, as before, to obtain the
following four-dimensional Einstein action. \beq S_4=\int
d^4x\sqrt{g_4}\lnk {M_4^2 \over 16 \pi} R_4
-\frac{1}{2}g^{\mu\nu}\partial_\mu\varphi\partial_\nu\varphi
-\frac{1}{2}g^{\mu\nu}\partial_\mu\chi\partial_\nu\chi
-\hat{U}(\varphi,\chi)\rnk, \eeq where the potential is given by
\beq \hat{U}(\varphi,\chi)=M^2_4M^2_5
e^{-\sqrt{\frac{2}{3}}\kappa\varphi}e^{-\frac{5\sqrt{3}}{6}\kappa\chi}
\lmk -\frac{bR_5^2}{2M_5^4} +
\frac{e^{\frac{\sqrt{3}}{2}\kappa\chi}-1}{64\pi
c}\frac{R_5}{M_5^2}\rmk, \label{usol} \eeq with
$\kappa\chi=\kappat\chih$ now.

If we had only the curvature-squared term in the original action
(\ref{5dqq}), we would find the potential
\beq
\hat{U}(\varphi,\chi)=M^2_4M^2_5
e^{-\sqrt{\frac{2}{3}}\kappa\varphi}e^{-\frac{5\sqrt{3}}{6}\kappa\chi}
\lmk e^{\frac{\sqrt{3}}{2}\kappa\chi} -1\rmk^2,
\eeq
which has no local maxima in the $\chi$ direction. It diverges as
$\hat{U}(\varphi,\chi)\propto e^{\frac{\sqrt{3}}{6}\kappa\chi}$ for
$\chi\longrightarrow \infty$ \cite{Maedar2}. On the other hand, in
the presence of both $R_5^2$ and $R_5^4$ terms in (\ref{5dqq}), the
latter term eventually dominates the former, and the potential
approaches zero asymptotically as in the pure quartic model
discussed above. Thus the curvature-squared term is expected to
increase $\chi_m$ and flatten the potential around the local
maximum.

\begin{figure}[tb]
 \begin{center}
\epsfig{file=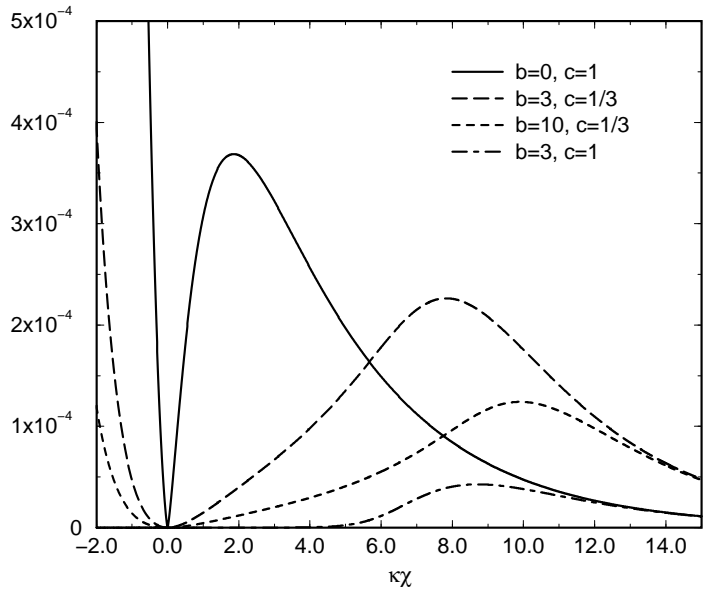,height=9cm}
\caption{\label{fig:fig2} {\it The potential from Eq.
(\protect\ref{usol}) for different values of the parameters $b$
and $c$ controlling the relative strengths of the $R^2$ and $R^4$
terms in the five-dimensional action.}} \ \end{center}
\end{figure}

Assuming that the dilaton has been stabilized as before, we have
numerically analyzed the potential $\hat{U}(0,\chi)$ for various
values of $b$ and $c$. As is seen in Fig.\ 2, the potential
becomes flatter as we increase $b$. At the same time, however, we
find that the height of the potential at the local maximum
decreases. As a result we find the maximal possible value of the
spectral index to be $n_s=0.722$ which is realized for $b \ga 10$
and $c \sim 1/3$. For example, taking $b=3$ and $c=1/3$, we find
$\chi_m=1.6M_4$, $\hat{U}(0,\chi_m)=2.3\times 10^{-4}M_4^2M_5^2$,
and $m^2/H_m^2=0.42$. The spectral index is given by $n_s=0.72$.
{}From $\delta T/T=10^{-5}$ on the angular scale probed by COBE, the
scale of inflation is determined as $H_m\simeq0.04M_5=6\times
10^{11}$ GeV or $M_5\simeq 2\times 10^{13}$ GeV. This is somewhat on
the small side, but we consider such an estimate relatively
encouraging, given the crudity of our model.

The scale problem concerning the $R^4$ inflation could also be
alleviated if consider the possible effects of the quantum
correction terms to the stress energy tensor as described in
(\ref{tmncor}) \cite{star}. Here we will concentrate on the term
proportional to $k_2$. Recall that the coefficient $k_2$ is
proportional to the number of degrees of freedom and has been
estimated to be $k_2 \sim O(100){\cal R}_5 M_5$. If we ignore the
$k_3$ term - which is actually just the variation of the $R^2$
term in the action, and has already been considered above -
and include our potential (\ref{upot}), (\ref{h'}) becomes \beq
H^2(1 - H^2/H'^2) = {\kappa^2 \over 3} U(\phi,\chi) \label{sol2}
\eeq where now $H'^2 = 360 \pi M_4^2 / k_2$. Near the maximum of
the potential at $\chi = \chi_m$, the potential is
$V(\chi_m)\simeq 10^{-3} M_4^2 M_5^2$. It is convenient to define
$\epsilon = (1 - {H^2 \over H'^2}) \simeq m^2/H^2$ where $m^2 \sim
3 \times 10^{-3} M_5^2 M_4^2$ is the curvature of the potential at
its maximum. Recall that our problem was related to the fact that
$m^2/H^2 \sim 1$.

Using (\ref{sol2}), we can now determine a consistent value for
$\epsilon$ and $m^2$ and therefore $H'$. We can then check whether
or not the resulting value of $k_2$ makes any sense with these
choices. To obtain the correct magnitude for density fluctuations
over the last $60$ e-foldings of inflation, we need $H \sim 10^{-5}
M_4 \epsilon e^{-20\epsilon}$. But recall that $H =
m/\sqrt{\epsilon}$, so that we require $M_5 \simeq 2 \times 10^{-3}
M_4 \epsilon^{3/2} e^{-20 \epsilon}$. We see that $M_5$ is
maximized when $\epsilon = 0.075$. In this case, we have $M_5 \sim
10^{-6} M_4 \sim 10^{13}$ GeV. As commented above, such an estimate
is rather too small, but we do not consider this too disastrous,
given the present level of sophistication of this class of model.

We should next determine whether or not we can obtain a value of
$k_2$ which is consistent with the desired values of $\epsilon$
and $M_5$. Substituting $k_2$ into $H'$, we find $H' \sim 3 M_4
/(M_5 {\cal R}_5)^{1/2}$. For our value of $\epsilon$, we need
$H'$ to be within 10\% of this value~\footnote{This is hardly fine
tuning by any standards.}, and we can solve for ${\cal R}_5^{-1}$.
The result is ${\cal R}_5^{-1} \sim 4 \times 10^{-15} M_5$, or
about 40 GeV. We note that some estimates of ${\cal R}_5$, based
on a version of the Scherk-Schwarz approach, could lead one to
expect ${\cal R}_5^{-1} \sim 100 M_5^3 /M_4^2 \sim 1$ TeV.
Therefore, we do not regard such a large value of ${\cal R}_5$ as
necessarily unmotivated. Finally, we can trivially check that
not only do we get sufficient inflation with an acceptable
magnitude for density fluctuations, but also find the spectral
index to be in agreement with the existing data. For $n_s$, using
$\epsilon = 3/40$, we have that $n_s = 1 - (1/20) = 0.95$, which
is a relatively encouraging result, well within the present
experimental uncertainties.

\section{Conclusion}

Many of the problems associated with inflation in string theory
can be traced to a rolling dilaton which precludes a (quasi) de
Sitter expansion. In this paper we have presented a model based on
higher-order curvature terms such as $R^4$ in the context of a
five-dimensional theory. 
This has some features in common with what
may arise in general compactifications of 
$M$ theory. 
In this context, the universe passes
through an extended five-dimensional phase (down from its initial
eleven-dimensional formulation) before it is seen as a four-
dimensional space-time with one relatively large extra spatial
dimension. We have made use of the conformal properties of the
theory to analyze inflation in the familiar four-dimensional Einstein
frame.
This has enabled us to identify the relevant degrees of freedom,
and consider their influence on cosmological dynamics.

We have found that a pure $R + R^4$ theory may provide a potential
suitable for topological inflation. However, in this theory the
spectral index for density fluctuations is too small, and the
magnitude of density fluctuations is too large, unless the
unification point is taken at an absurdly low energy scale. We
have considered several ways of repairing these problems. One
approach may be to include $R^2$ corrections in the action. Such
corrections are not known to arise in superstring theories, but may appear
if the supersymmetry breaking scale and the unification scale are
close to each other. They can soften the effective inflaton
potential further. Alternatively, a similar flattening of the
potential may arise as a consequence of a large number of particle
degrees of freedom, expected as the Kaluza-Klein modes 
associated with an extended
compact spatial dimension, which could be produced in the course
of cosmological evolution. Particle production can likewise raise
the spectral index while keeping the magnitude of fluctuations
reasonable. This result calls for a rather low unification scale
of $10^{13}$ GeV and a scale ${\cal R}_5^{-1} \sim 1$ TeV for the
fifth dimension. Given the crudity of this model, these results must
be seen as encouraging.

To conclude, we note that our arguments do not depend strongly on
the specific form of higher-order corrections in the effective
action. We only need some higher-order curvature corrections which
produce a local extremum of the effective inflaton potential. It
seems plausible that such terms may play a significant role in
some region of a highly curved early universe. The usual fine
tuning of the coefficients of these terms, which was detrimental
for the Starobinsky model, could be avoided here because of the
dilaton v.e.v.. In turn, the quadratic corrections may be
employed to flatten the potential in order to produce a
satisfactory density perturbation spectrum, while not being
overconstrained thanks to the fact that inflation is driven by
higher corrections. Finally, the graceful exit problem is very
easy to solve, due to the inherent instability of the solutions.
The inflaton rolls down to the minimum, slowly at first but
accelerating along the way. In light of this, we take this model
to be a reasonable candidate for standard quasi-exponential
inflation at scales close to the unification scale, which might
offer some clues how to embed inflation in string theory.

\vspace{1cm}
{\bf Acknowledgments}

We would like to thank I. Antoniadis for helpful discussions.
N.K. was supported in part by NSF Grant PHY-9219345,
K.A.O. was supported in part by DOE
grant DE--FG02--94ER--40823 and J.Y. was supported in part by
the Japanese Grant in Aid of Science Research Fund of the
Monbusho No. 09740334  and ``Priority Area: Supersymmetry and Unified
Theory of Elementary Particles (No.\ 707).

\end{document}